\documentclass[11pt]{article}
\usepackage{a4}

\usepackage{amsfonts}
\usepackage{amsmath}
\usepackage{amssymb}

\newcommand{\bea}{\begin{eqnarray}}
\newcommand{\eea}{\end{eqnarray}}
\newcommand{\be}{\begin{equation}}
\newcommand{\ee}{\end{equation}}
\newcommand{\pa}{\partial}
\newcommand{\nn}{\nonumber}

\newcommand{\e}{\epsilon}

\newcommand{\dmu}{{\dot \mu}}
\newcommand{\dnu}{{\dot \nu}}
\newcommand{\drho}{{\dot \rho}}

\newcommand{\dlambda}{{\dot \lambda}}

\newcommand{\hmu}{{\hat \mu}}
\newcommand{\hnu}{{\hat \nu}}
\newcommand{\hrho}{{\hat \rho}}
\newcommand{\hsigma}{{\hat \sigma}}

\newcommand{\bc}{\bigcirc}

\makeatletter

\@addtoreset{equation}{section}
\makeatother

\def\href#1#2{#2}

\begin{document}

\begin{titlepage}

\begin{center}

\hfill 
\vskip 1.1in

{\bf \Large String solitons in the M5-brane worldvolume with}\\[2mm]
{\bf \Large a Nambu-Poisson structure and Seiberg-Witten map}

\vskip .7in
{\sc 
FURUUCHI Kazuyuki$\,{}^{a,}$\footnote{furuuchi@phys.cts.nthu.edu.tw} and\ 
TAKIMI Tomohisa$^{b,}$\footnote{takimi@phys.ntu.edu.tw}}\\
\vskip 10mm
{\sl
${}^a$%
National Center for Theoretical Sciences, \\
National Tsing-Hua University, Hsinchu 30013, Taiwan,
R.O.C.}\\
\vskip 2mm
{\sl
${}^b$%
Department of Physics, 
National Taiwan University, \\
Taipei 10617, Taiwan,
R.O.C.}\\

\vspace{11pt}
\end{center}
\begin{abstract}
We analyze BPS equations for string-like configurations 
derived from the M5-brane worldvolume action
with a Nambu-Poisson structure
constructed in 
Ref.\cite{Ho:2008nn,Ho:2008ve}.
We solve the BPS equations 
up to the first order in the parameter $g$
which characterizes the strength of the Nambu-Poisson bracket.
We compare our solutions to 
previously constructed BPS string solitons 
in the conventional description of M5-brane  
in a constant three-form background
via Seiberg-Witten map,
and find agreement.
\end{abstract}

\end{titlepage}

\setcounter{footnote}{0}

\section{Introduction}

M-theory \cite{Witten:1995ex}
has been a powerful guide in the study of non-perturbative aspects of string theory.
However, its microscopic formulation is still lacking.
M-theory branes are important building blocks
of M-theory and deeper understanding of them
will be crucial for making progress regarding this issue.

Recently a model for
multiple M-theory membranes 
based on Lie 3-algebra proposed in 
Ref.\cite{Bagger:2006sk,Bagger:2007jr,Gustavsson:2007vu}
has been intensively studied.
The model has several promising features
for a correct description
of multiple M-theory membranes at low energy.
On the other hand,
its relations to the space-time covariant formulation of M-theory branes
are not fully clarified yet.\footnote{%
See \cite{Park:2008qe,Furuuchi:2008ki,Bandos:2008fr,Furuuchi:2009ax} 
for investigations on this issue.}
In Ref.\cite{Ho:2008nn,Ho:2008ve}
a new M5-brane action was constructed
from the multiple membrane action
by choosing a Nambu-Poisson algebra as Lie 3-algebra.
It was proposed that
this new action 
may be mapped to more conventional
(``ordinary" in the following)
description of
M5-brane \cite{Howe:1996yn,Howe:1997fb,%
Pasti:1997gx,Bandos:1997ui,Aganagic:1997zq,Bandos:1997gm}
in a constant three-form background,\footnote{%
Note that space-time covariance is broken
only by fixing the three-form background in this case.}
in analogy with 
D-branes in a constant B-field background
which has non-commutative and commutative descriptions
related via Seiberg-Witten map \cite{Seiberg:1999vs}.

In this paper we study BPS string solitons in 
the M5-brane worldvolume with a Nambu-Poisson structure.
These configurations
describe an
M2-brane ending on an M5-brane.
Such BPS string solitons were first
constructed in Ref.\cite{Howe:1997ue}
in the conventional description of M5-brane,
and they were generalized to the case 
with a constant three-form background
in Ref.\cite{Michishita:2000hu,Youm:2000kr}.
From M2-brane worldvolume action,
this type of configurations 
with a constant three-form field background has been
studied in Ref.\cite{Bergshoeff:2000jn,Kawamoto:2000zt}.
More recently, it was studied
from the multiple membrane action in Ref.\cite{Chu:2009iv}
through the deformed 
Basu-Harvey equation \cite{Basu:2004ed},
and their work may be complementary to present work.
We solve the BPS equations
in the first order in the parameter $g$
which characterizes the strength of the 
Nambu-Poisson bracket.
We compare our solutions
with the previously constructed BPS string solitons
in the ordinary description of M5-brane in constant three-form flux
via the Seiberg-Witten map \cite{Ho:2008ve,Seiberg:1999vs},
and find nice agreement.

\section{String solitons in
the M5-brane worldvolume with a Nambu-Poisson structure}

\subsection{Supersymmetry transformation in the M5-brane action}

In this subsection we 
review the supersymmetry transformation
in the M5-brane worldvolume action 
with a Nambu-Poisson structure
constructed in Ref.\cite{Ho:2008nn,Ho:2008ve}
to fix our notation and prepare for the study
of BPS equations in the subsequent subsections.
The detail of the construction of the M5-brane action
can be found in Ref.\cite{Ho:2008ve}.
We will follow the notation of Ref.\cite{Ho:2008ve}
except that we omit ``\,'\," from the six dimensional variables
and some obvious modifications in the numbering of coordinates.
In this model the supersymmetry transformation of the 
fermionic field $\Psi$ is given as follows:
\begin{eqnarray}
 \label{SUSYPsi}
\delta \Psi
&=&{\cal D}_\mu X^i\Gamma^\mu\Gamma^i\epsilon
+{\cal D}_{\dot\mu}X^i\Gamma^{\dot\mu}\Gamma^i\epsilon
\nonumber\\&&
-\frac{1}{2}
{\cal H}_{\mu\dot\nu\dot\rho}
\Gamma^\mu\Gamma^{\dot\nu\dot\rho}\epsilon
-{\cal H}_{345}
\Gamma_{345}\epsilon
\nonumber\\&&
-\frac{g^2}{2}\{X^{\dot\mu},X^i,X^j\}
\Gamma^{\dot\mu}\Gamma^{ij}\epsilon
+\frac{g^2}{6}\{X^i,X^j,X^k\}
\Gamma^{ijk}\Gamma^{345}\epsilon ,
\end{eqnarray}
where fields live
on the six dimensional M5-brane worldvolume
parametrized by
$x^\mu$ ($\mu = 0,1,2$) and
$y^\dmu$ ($\dmu = 3,4,5$).
$X^i$'s ($i=6,\cdots,10$) are scalar fields
which describe embedding of the M5-brane in the transverse space.
$\Gamma$'s are eleven dimensional Gamma matrices.
The metric on the M5-brane is mostly plus,
diag$(-1,1,1,\cdots,1)$.
The fermionic shift symmetry 
has already been taken into account
so that the configuration that
all the fields vanish is invariant under
the supersymmetry
(see Section 6 of Ref.\cite{Ho:2008ve} for more detail).
The chirality of the fermion and supersymmetry parameters are 
chosen as follows:
\bea
\Gamma^{012345} \Psi = -\Psi , \qquad
\Gamma^{012345} \e = \e .
\eea
$\{ *, *, * \}$ denotes
the Nambu-Poisson bracket 
which we choose to be the one on $\mathbb{R}^3$:
\bea
 \label{NP}
\{f,g,h\}
=
\e^{\dmu\dnu\drho} 
\frac{\pa f}{\pa y^\dmu} \frac{\pa g}{\pa y^\dnu} \frac{\pa h}{\pa y^\drho}  ,
\eea
where $\e^{\dmu\dnu\drho}$ is a totally anti-symmetric
tensor on $\mathbb{R}^3$ with $\e^{345} = 1$.
$X^{\dmu}$ is given by
\bea
X^{\dmu} = \frac{y^{\dmu}}{g} + b^{\dmu}, \qquad
b^{\dmu} = \frac{1}{2}\e^{\dmu\dnu\drho}b_{\dnu\drho}.
\eea
The covariant derivatives 
in the directions
$\mu = 0,1,2$ are given as
\bea
{\cal D}_\mu\varphi
&\equiv&D_\mu\varphi
=\partial_\mu\varphi
-g\{b_{\mu\dot\nu},y^{\dot\nu},\varphi\},\label{dmu}
\eea
and those in the directions $\dmu = 3,4,5$ are given by
\begin{eqnarray}
{\cal D}_{\dot\mu}\varphi
&\equiv&\frac{g^2}{2}\epsilon_{\dot\mu\dot\nu\dot\rho}
\{X^{\dot\nu},X^{\dot\rho},\varphi\}
\nonumber\\
&=&
\partial_{\dot\mu}\varphi
+g(
\partial_{\dot\lambda}b^{\dot\lambda}\partial_{\dot\mu}\varphi
-\partial_{\dot\mu}b^{\dot\lambda}\partial_{\dot\lambda}\varphi
)
+\frac{g^2}{2}\epsilon_{\dot\mu\dot\nu\dot\rho}
\{b^{\dot\nu},b^{\dot\rho},\varphi\}.\label{ddotmu}
\end{eqnarray}
Here, $\varphi$ collectively represents ``covariant" fields
$X^i$ and $\Psi$.
The field strength of the anti-symmetric tensor field is given by
\begin{eqnarray}
{\cal H}_{\lambda\dot\mu\dot\nu}
&=&\epsilon_{\dot\mu\dot\nu\dot\lambda}{\cal D}_\lambda X^{\dot\lambda}
\nonumber\\
&=&H_{\lambda\dot\mu\dot\nu}
-g\epsilon^{\dot\sigma\dot\tau\dot\rho}
(\partial_{\dot\sigma}b_{\lambda\dot\tau})
\partial_{\dot\rho}b_{\dot\mu\dot\nu},\label{h12def}\\
{\cal H}_{345}
&=&g^2\{X^{3},X^{4},X^{5}\}-\frac{1}{g}
=\frac{1}{g}(V-1)
\nonumber\\
&=&H_{345}
+\frac{g}{2}
(\partial_{\dot\mu}b^{\dot\mu}\partial_{\dot\nu}b^{\dot\nu}
-\partial_{\dot\mu}b^{\dot\nu}\partial_{\dot\nu}b^{\dot\mu})
+g^2\{b^{3},b^{4},b^{5}\}, 
\label{h30def}
\end{eqnarray}
where $V$ is the ``induced volume''
\begin{equation}
V=g^3\{X^{3},X^{4},X^{5}\},
\end{equation}
and $H$ is the linear part of the field strength
\begin{eqnarray}
H_{\lambda\dot\mu\dot\nu}
&=&
\partial_{\lambda}b_{\dot\mu\dot\nu}
-\partial_{\dot\mu}b_{\lambda\dot\nu}
+\partial_{\dot\nu}b_{\lambda\dot\mu},\\
H_{\dot\lambda\dot\mu\dot\nu}
&=&
\partial_{\dot\lambda}b_{\dot\mu\dot\nu}
+\partial_{\dot\mu}b_{\dot\nu\dot\lambda}
+\partial_{\dot\nu}b_{\dot\lambda\dot\mu}.
\end{eqnarray}

\subsection{BPS equations for string-like configurations}

The type of brane configurations we will study is as follows:
\begin{center}
\begin{tabular}{cccccccccccc}
          & 0 & 1 & 2 & 3 & 4 & 5 & 6 &7&8&9&10\\
M5        &$\bc$&$\bc$&$\bc$&$\bc$&$\bc$&$\bc$& $-$  & $-$ & $-$ & $-$ & $-$  \\
M2 soliton&$\bc$&$\bc$& $-$  & $-$ &  $-$ & $-$  & $\bc$ & $-$ & $-$ & $-$ & $-$ 
\end{tabular}
\end{center}
where $\bc$
denotes the direction the brane extends and $-$ denotes
the direction the brane localizes.
The M2-brane ending on the M5-brane 
appears as a string in the M5-brane worldvolume
extending in the 1-st direction.
Note that the Nambu-Poisson structure is in the $345$ directions.
We will study the configurations which preserve
half of the supersymmetry parametrized by
\bea
\Gamma^{016} \e = \mp \e .
\eea
From Eq.(\ref{SUSYPsi}) we observe that
the supersymmetry transformation parametrized 
by above $\e$ is preserved 
when the following BPS equations are satisfied: 
\bea
 \label{BPS}
{\cal D}_\hmu X^6 \pm
\frac{1}{6}
\e_{\hmu}{}^{\hnu \hrho \hsigma} 
{\cal H}_{\hnu \hrho \hsigma} = 0,
\eea
and other fields set to zero,
where $\hmu,\hnu = 2,\cdots,5$.
$\e^{\hmu\hnu \hrho \hsigma}$
is a totally anti-symmetric tensor with
$\e^{2345}=1$.

\subsection{BPS equations and solutions at order $g^0$}

We construct the solutions to the BPS equations (\ref{BPS})
by expansions in $g$:
\bea
X^6 \equiv \Phi 
&=& \Phi_{(0)} + g \Phi_{(1)} + g^2 \Phi_{(2)} + {\cal O}(g^3), \nn \\
b_{\mu\dnu}  &=& b_{\mu\dnu(0)} + g b_{\mu\dnu(1)} + g^2 b_{\mu\dnu(2)}+ {\cal O}(g^3), \nn \\
b_{\dmu\dnu} &=& b_{\dmu\dnu(0)} + g b_{\dmu\dnu(1)} + g^2 b_{\dmu\dnu(2)} + {\cal O}(g^3) .
\eea
At order $g^0$, the BPS equation (\ref{BPS}) becomes
\bea
\pa_{\hmu} \Phi_{(0)} 
\pm
\frac{1}{6}
\e_{\hmu}{}^{\hnu \hrho \hsigma} 
{H}_{\hnu \hrho \hsigma(0)} = 0.
\eea
From the condition that $H_{(0)}$ can be written 
as $H_{(0)} = db_{(0)}$ in an open patch,
i.e. from the condition $dH_{(0)} =0$,
we obtain the condition
\bea
 \label{Phi0}
\Box \Phi_{(0)} = 0,
\eea
where $\Box \equiv \delta^{\hmu\hnu}\pa_{\hmu}\pa_{\hnu}$.
We consider delta-function source at the origin,
like in the case of Dirac monopole,
and Eq.(\ref{Phi0}) is not satisfied globally.
Thus we have
\bea
\Phi_{(0)} = \frac{m}{r^2},
\eea
where $r^2 = \sum_{\hmu=2}^{5} (x_{\hmu})^2$
and 
\bea
m = \frac{k}{(2\pi)^{3/2} \sqrt{T_6}},
\eea
where 
the integer $k$ is a topological charge of the solution, 
and $T_6$ is the tension of the 
M5-brane with the Nambu-Poisson structure.\footnote{%
We follow the notation of Ref.\cite{Ho:2008ve}.
See section 7 of the reference for the Dirac quantization condition.}
Corresponding tensor field configurations are given by
\bea
b^{\dmu}_{(0)} = \mp \frac{m x_{\dmu}}{a^3} A , 
\eea
where
\bea
\label{defA}
A = \pm \frac{\pi}{2} 
+ \tan^{-1} \left( \frac{x_2}{a} \right) + \frac{a x_2}{r^2} ,
\eea
with
\bea
a^2 = x^2_3 + x^2_4 + x^2_5 .
\eea
The solution for the tensor field
has been studied in Ref.\cite{Nepomechie:1984wu}.
Note that we have chosen the gauge
$b_{23}=b_{24}=b_{25}=0$ which simplifies our analysis.
The choice of $\pm$ in (\ref{defA})
corresponds to the choice of the direction of the
Dirac string.
At order $g^0$ 
the Dirac string is not physical.
The $g$ expansion is not 
a good expansion for studying the 
fate of the Dirac string,
since $g$ is associated
with the Nambu-Poisson bracket
which has three derivatives,
and it follows that it is actually the
expansion in ${gm}/ a^3$.
One can deduce it from  
the mass dimension counting and
the explicit form of the 
zero-th order solution.
Table \ref{massd} summarizes the mass dimension
of the relevant fields and parameters in our convention
for readers' convenience.
Such expansion is not appropriate for $a^3 \lesssim gm$.
Therefore, in the rest of the paper
we are satisfied with 
that the Dirac string is a gauge artifact
at order $g^0$ and
do not worry too much about the Dirac string.
In the case of monopoles
in non-commutative space,
it has been shown 
that the Dirac string becomes physical and 
smooth due to the effect non-perturbative in 
the non-commutative parameter \cite{Gross:2000wc}.

\begin{table}[htbp]
\begin{center}
\begin{tabular}{|c|c|c|c|c|c|c|}
\hline
               & $y^\dmu$ & $X^i$ ($\Phi \equiv X^6$) & $b_{\mu\dnu}$, $b_{\dmu\dnu}$ & $g$ & $m$ & $T_6$  \\ \hline
mass dimension & $-1$     & $-1$  & $-1$ & $0$ & $-3$ & $6$  \\
\hline
\end{tabular}
\caption{Mass dimension of the relevant fields and parameters.}
\label{massd}
\end{center}
\end{table}

\subsection{BPS equations and solutions at order $g$}

Now we move on to the order $g$
solutions.
We should solve
\begin{eqnarray}
0 &=& \pa_{2} \Phi_{(1)} 
\pm 
\left(
H_{345(1)} + \frac{1}{2}(\pa_{\dmu}b^{\dmu}_{(0)}\pa_{\dnu}b^{\dnu}_{(0)}
-\pa_{\dmu}b^{\dnu}_{(0)}\pa_{\dnu}b^{\dmu}_{(0)}) 
\right),
\\
0 &=& \pa_{\dlambda} \Phi_{(1)} + 
(\pa_{\dmu}b^{\dmu}_{(0)}\pa_{\dlambda}\Phi_{(0)}
-\pa_{\dlambda}b^{\dnu}_{(0)}\pa_{\dnu}\Phi_{(0)})
\pm \frac{1}{2} \e_{\dlambda}{}^{2\dmu\dnu}H_{2\dmu\dnu(1)}. 
\end{eqnarray}
As in the previous subsection,
we solve the condition
$dH_{(1)}=0$.
This condition reduces to
\bea
 \label{lap1}
\Box \Phi_{(1)} =  
\pm m^2
\left(
\frac{16x_2}{r^8} + \frac{32A}{ar^6}
\right).
\eea
We found
\bea
 \label{NPsol}
\Phi_{(1)} 
= 
\pm
m^2
\left(
\frac{2x_2}{r^6} + \frac{2A}{ar^4} 
\right) ,
\eea
solves Eq.(\ref{lap1}) while
it does not modify 
the boundary conditions on $\Phi$ 
at $r \rightarrow \infty$.
Note that $\Phi$ is not gauge invariant.
(See Ref.\cite{Ho:2008ve} for gauge transformation laws in
the M5-brane worldvolume action with the Nambu-Poisson structure.)

\subsection{Seiberg-Witten map}

Seiberg-Witten map 
was first found as a map
between non-commutative description
and commutative description of D-branes
in a constant B-field background \cite{Seiberg:1999vs}.
In Ref.\cite{Ho:2008ve} it
was generalized to a map
between
description by the M5-brane with the Nambu-Poisson structure
and description by the ordinary M5-brane in a
constant three-form background.
As a first step, we
study the Seiberg-Witten map for
the scalar field. 
Only in this subsection, 
we denote the scalar field
in the Nambu-Poisson description as
$\hat{\Phi}$, and the corresponding field
in the ordinary description as $\Phi$.

BPS string solitons
in M5-brane in constant three-form flux
have been constructed in Ref.\cite{Michishita:2000hu,Youm:2000kr}.
The scalar configuration $\check{\Phi}$ is given by
\bea
\check{\Phi} =
\frac{m}{\check{r}^2} \pm \tan \theta \, \check{x}_2  ,
\eea
where $\theta$ is related to the 
background three-form field as
$H_{345}^{(bg)} =  - \tan \theta$.\footnote{%
The solution looks like the one for
linearized M5-brane action, but
actually it solves the equation of motion of the
non-linear M5-brane action in the ordinary description \cite{Michishita:2000hu,Youm:2000kr}.
Our convention differs from that in Ref.\cite{Michishita:2000hu,Youm:2000kr} 
by a factor of $\frac{1}{4}$.}
Here $\check{x}_{\dmu}=x_\dmu$ ($\dmu = 3,4,5$),
but for $\check{x}_2$
we need to make a rotation in the coordinate
and the field before applying the 
Seiberg-Witten map \cite{Hashimoto:2000mt,Mateos:2000qq,%
Moriyama:2000mm,Hashimoto:2000cf}:
\bea
\left(
\begin{array}{c}
{\Phi}\\
{x}_2
\end{array}
\right)
=
\left(
\begin{array}{cc}
\cos \phi & -\sin \phi \\
\sin \phi & \cos \phi
\end{array}
\right)
\left(
\begin{array}{c}
\check{\Phi}\\
\check{x}_2
\end{array}
\right)  ,
\eea
where we choose $\phi$
so that 
the term linear in $x_2$ does
not appear in $\Phi$.
For $|\theta | \ll 1$, $\phi = \pm \theta + {\cal O}(\theta^2)$.

The Seiberg-Witten map for the scalar field is given in 
Ref.\cite{Ho:2008ve} up to order $g$:
\bea
\hat{\Phi}
=
\Phi+
g b^{\dmu} \pa_{\dmu} \Phi + {\cal O}(g) .
\eea
Up to the first order in $g$ and $\theta$,
we obtain
\bea
\hat{\Phi} = \frac{m}{r^2}
\pm 2m^2
\left(
\frac{\theta x_2}{r^6} + \frac{gA}{a r^4}
\right)
+ \mbox{(higher order terms in $g$ and $\theta$)}.
\eea
This coincides with 
our solution (\ref{NPsol})
if 
\bea
 \label{gt}
g = \theta + {\cal O}(\theta^2).
\eea
Eq.(\ref{gt}) is consistent with the result of Ref.\cite{Ho:2008ve}.

\section{Summary and future directions}

In this paper we obtained solutions of the 
BPS equations for string-like configurations 
derived from the M5-brane worldvolume action 
with the Nambu-Poisson structure 
constructed in Ref.\cite{Ho:2008nn,Ho:2008ve}
up to the first order in $g$.
After the Seiberg-Witten map our solutions agreed with
the BPS string solitons in the ordinary description of M5-brane.
This result motivates  
more thorough study of
Seiberg-Witten map
between 
the M5-brane worldvolume action with the Nambu-Poisson structure
and 
the ordinary M5-brane worldvolume action 
in constant three-form flux.
In the case of D-branes in a constant B-field background,
it has been argued 
(and explicitly checked for the first few terms 
in the expansion in the slowly varying field strength $F$
of the ordinary gauge field)
that 
the non-commutative $\hat{F}^2$ action coincides
with the ordinary DBI action
in the zero slope limit,
up to total derivative terms and an additive constant \cite{Seiberg:1999vs}.
The M5-brane action with the Nambu-Poisson structure,
with the ${\cal H}^2$ term in it being in parallel with the $\hat{F}^2$ term above,
may similarly coincide
with the ordinary (DBI-type) M5-brane action
in an appropriate limit of the M2-brane tension 
and the background three-form flux.
But this needs to be checked by further investigation.\footnote{%
It would be worthwhile to mention
that in the case of D-brane in a constant B-field background,
the description by
the Poisson bracket may not be just an approximation
of the description by the Moyal product, 
but it can be another description 
related to commutative or non-commutative descriptions
through Seiberg-Witten (type) maps
\cite{Ishibashi:1999vi,Okuyama:1999ig}.
Similar story may hold in
the case of M5-brane in constant three-form flux.}
To achieve this goal, we first need to understand
how to take the appropriate limit.
This might not be as simple as in the case 
of D-branes in a constant B-field background
which can be studied using the open string worldsheet free CFT, 
due to the interacting nature of the membrane worldvolume theory
(see Ref.\cite{Bergshoeff:2000jn,Kawamoto:2000zt}
for earlier studies).
But the investigation through
the relation between M-theory and type IIA string theory
along the line of Ref.\cite{Ho:2008ve}
may be of help to understand this issue. 
We also need to understand 
how to connect the apparently different treatments of
the self-dual two-form between the two descriptions of 
the M5-brane
in constant three-form flux.\footnote{%
After we added these lines in our draft
in response to the referee's comment
while we were preparing the final revision of this paper,
Ref.\cite{Pasti:2009xc} appeared which 
made an interesting progress in this direction.}

Our analysis was restricted to 
the expansion in the parameter $g$.
Such expansion is not suitable for studying the structure
near the Dirac string.
In the case of solitons/instantons 
in non-commutative space,
techniques to obtain exact solutions
have been developed
by expressing functions on non-commutative space
with operators acting on the Hilbert space of harmonic oscillators
\cite{Nekrasov:1998ss,Furuuchi:1999kv,Furuuchi:2000vx,Furuuchi:2000vc,%
Gopakumar:2000zd,Gross:2000wc,%
Polychronakos:2000zm,%
Gross:2000ph,%
Bak:2000ac,Aganagic:2000mh,Harvey:2000jb,
Furuuchi:2000dx%
}.
In these cases solutions are smooth due to the effect
non-perturbative in the non-commutative parameter.
To construct solutions 
on the M5-brane with the Nambu-Poisson structure
in a similar way,
we would first need to understand what is the appropriate
``quantization" of the Nambu-Poisson bracket.
For this purpose, investigations in 
Ref.\cite{Chu:2009iv,Ho:2007vk} seem very suggestive.
It will be very interesting to study
solitons on manifolds 
with a quantum Nambu-Poisson structure
from the M-theory point of view.

\section*{Acknowledgments}
The authors thank 
Chien-Ho Chen,
Takayuki Hirayama,
Shoichi Kawamoto,
Sheng-Yu Darren Shih
and especially Pei-Ming Ho 
for explanations on
their works as well as helpful discussions.
They also thank Dan Tomino for useful discussions.
T.T. is grateful to the members of
the Harish-Chandra Research Institute, Allahabad, India
for kind hospitality during his visit in January to February, 2009.
He is also thankful for the support for the trip
from the Strings Focus Group, NCTS, Taiwan.
This work is supported in part
by 
National Science Council of Taiwan
under grant No. NSC 97-2119-M-002-001 (F.K.,T.T.)
and No. NSC 97-2811-M-002-125 (T.T).

\bibliography{M2onM5refs}
\bibliographystyle{utphys}

\end{document}